\documentstyle[12pt,epsf]{article}
\begin{document}
\begin{center}
{\large \bf Lifetimes of doubly charmed baryons:\\ $\Xi_{cc}^{+}$ and
$\Xi_{cc}^{++}$. 
}\\
\vspace*{5mm}
V.V. Kiselev, A.K. Likhoded, A.I. Onishchenko\\
{\sf State Research Center of Russia "Institute for High Energy Physics"} \\
{\it Protvino, Moscow region, 142284 Russia}\\
Fax: +7-095-2302337\\
E-mail: kiselev@mx.ihep.su
\end{center}
\abstract{
We have performed a detailed investigation of total lifetimes for the
$\Xi_{cc}^{++}$ and $\Xi_{cc}^{+}$ baryons in the framework of operator
product expansion over the inverse mass of charmed quark, whereas,
to estimate matrix elements of operators obtained in OPE, some approximations
of nonrelativistic QCD are used. This approach allows one to take into account
the corrections to the spectator decays of $c$-quarks, which reflect the fact,
that these quarks are bound, as well as the contributions, connected to the
effects of both the Pauli interference for the $\Xi_{cc}^{++}$-baryon and the
weak scattering for the $\Xi_{cc}^{+}$-baryon. The realization of such program
leads to the following estimates for the total lifetimes of doubly charmed
baryons: $\tau_{\Xi_{cc}^{++}}=0.43\pm 0.1~ps$ and $\tau_{\Xi_{cc}^{+}}=0.11\pm
0.01~ps$.
}\\

\newpage
\section{Introduction}
A study on weak decays of doubly charmed baryons is of great interest
because of two reasons. The first one is connected to the investigation on
the basic properties of weak interactions at the fundamental level, including
the precise determination of CKM matrix parameters. The second reason is
related to the possibility to explore QCD as it is provided by the systems
containing the heavy quarks. In the limit of a large scale given by the heavy
quark mass, some aspects in the dynamics of strong interactions become simpler
and one gets a possibility to draw definite model-independent predictions. Of
course, both these topics appear in the analysis of weak decays for the doubly
charmed baryons, whose dynamics is determined by an interplay between the
strong and weak interactions. That is why these baryons are the attractive and
reasonable subjects for the theoretical and experimental consideration.

The doubly charmed $\Xi_{cc}^{(*)}$-baryon
represents an absolutely  new type of
objects in comparison with the ordinary baryons containing
light quarks only. The basic state of  such baryon  is analogous
to a $(\bar Q q)$-meson, which contains a single heavy antiquark
$\bar Q$ and a light quark $q$. In the doubly heavy baryon the role
of heavy antiquark is played by the $(cc)$-diquark, which is in
antitriplet color-state. It has a small size in comparison
with the scale of the light quark confinement. Nevertheless the spectrum of
$(ccq)$-system states has to  differ essentially from the heavy meson
spectra, because the composed $(cc)$-diquark has a set of the excited
states (for example, $2S$ and $2P$) in contrast to the heavy quark.
The energy of diquark excitation is twice less than the excitation
energy of light quark bound with the diquark. So, the
representation on the compact diquark can be straightforwardly connected to
the level structure of  doubly heavy baryon.

Naive estimates for the lifetimes of doubly charmed baryons were done
by the authors slightly early \cite{1}. A simple consideration of
quark diagrams shows, that in the decay of  $\Xi_{cc}^{++}$-baryons,
the Pauli interference for the decay products of charmed quark and the valent
quark in the initial state takes place in an analogous way to the $D^+$-meson
decay. In the decay of $\Xi_{cc}^+$, the exchange by the $W$-boson  between
the valence quarks plays an important role like in the decay of $D^0$. These
speculations and the presence of two charmed quarks in the initial
state result in the following estimates for the lifetimes:
\begin{eqnarray}
\tau(\Xi_{cc}^{++})&\approx & \frac{1}{2} \tau(D^+)\simeq 0.53\; {\rm
ps,}
\nonumber \\
\tau(\Xi_{cc}^{+})&\approx & \frac{1}{2} \tau(D^0)\simeq 0.21\;{\rm
ps.}
\nonumber
\end{eqnarray}
   
In this work we discuss the systematic approach to the evaluation of total
lifetimes for the doubly charmed baryons on the basis of both the optical
theorem
for the inclusive decay width and the operator product expansion (OPE) for the
transition currents in accordance with the consequent nonrelativistic expansion
of hadronic matrix elements derived in OPE. Using OPE at the first step,
we exploit the fact, that, due to the presence of heavy quarks in the initial
state, the energy release in the decay of both quarks is large enough in
comparison with the binding energy in the state. Thus, we can use the expansion
over the ratio of these scales. Technically, this step repeats an analogous
procedure for the inclusive decays of heavy-light mesons as it was reviewed in
\cite{2}. Exploring the nonrelativistic expansion of hadronic matrix elements
at the second step, we use the approximation of nonrelativistic QCD \cite{3,4},
which allows one to reduce the evaluation of matrix elements for the full QCD
operators, corresponding to the interaction of heavy quarks inside the diquark,
to the expansion in powers of $\frac{p_c}{m_c}$, where $p_c =m_cv_c\sim 1~GeV$
is a typical momentum of the heavy quark inside the baryon. The same procedure
for the matrix elements, determined by the strong interaction of heavy quarks
with the light quark, leads to the expansion in powers of
$\frac{\Lambda_{QCD}}{m_c}$.

This way, taking into account the radiation of hard gluons in these decays,
leads to the expansion in powers of $\alpha_s$, $v_c = \frac{p}{m_c}$ and
$\frac{\Lambda_{QCD}}{m_c}$. It is worth to note, that this expansion would be
well defined, provided the expansion parameters to be small. In the $c\to
su\bar d$ transition, the ratio of typical momentum for the heavy quark
inside the hadron to the value of energy, released in the decay, is not so
small. We would like also to stress the important roles, played by both the
Pauli interference and the weak scattering, suppressed as $\frac{1}{m_c^3}$
with respect to the leading spectator contribution, but the former ones are
enhanced by a numerical factor, caused by the ratio of two-particle   
and three-particle phase spaces \cite{5}. Numerical estimates show that the
value of these contributions is considerably large, and it is of the order of
$40 -140\%$. These effects take place in the different baryons,
$\Xi_{cc}^{++}$ and $\Xi_{cc}^{+}$, and, thus, they enhance the difference of
lifetimes for these baryons. The final result for the total lifetimes of doubly
charmed baryons is the following: 
\begin{eqnarray}
\tau_{\Xi_{cc}^{++}} &=& 0.43\pm 0.1~ps, \nonumber\\
\tau_{\Xi_{cc}^{+}} &=& 0.11\pm 0.01~ps. \nonumber
\end{eqnarray}

\section{Operator product expansion}
Now let us start the description of our approach for the calculation of total
lifetimes for the doubly charmed baryons. The optical theorem, taking into
account the integral quark-hadron duality, allows us to relate the total decay
width of the heavy quark with the imaginary part of its forward scattering
amplitude. This relationship, applied to the $\Xi_{cc}^{(*)}$-baryon total
decay width $\Gamma_{\Xi_{cc}^{(*)}}$, can be written down as:
\begin{equation}
\Gamma_{\Xi_{cc}^{(*)}} =
\frac{1}{2M_{\Xi_{cc}^{(*)}}}\langle\Xi_{cc}^{(*)}|{\cal T}
|\Xi_{cc}^{(*)}\rangle ,
\label{1}
\end{equation}
where the $\Xi_{cc}^{(*)}$ state in (\ref{1}) has the ordinary relativistic
normalization, $\langle \Xi_{cc}^{(*)}|\Xi_{cc}^{(*)}\rangle  = 2EV$, and  the
transition operator ${\cal T}$ is determined by the expression
\begin{equation}
{\cal T} = \Im m\int d^4x~\{{\hat T}H_{eff}(x)H_{eff}(0)\},
\end{equation}
where $H_{eff}$ is the standard effective hamiltonian, describing the low
energy interactions of initial quarks with the decays products, so that
\begin{equation}
H_{eff} = \frac{G_F}{2\sqrt 2}V_{uq_1}V_{cq_1}^{*}[C_{+}(\mu)O_{+} +
C_{-}(\mu)O_{-}] + h.c. 
\end{equation}
where 
$$
O_{\pm} = [\bar q_{1\alpha}\gamma_{\nu}(1-\gamma_5)c_{\beta}][\bar
u_{\gamma}\gamma^{\nu}(1-\gamma_5)q_{2\delta}](\delta_{\alpha\beta}\delta_{
\gamma\delta}\pm\delta_{\alpha\delta}\delta_{\gamma\beta}),
$$
and
$$
C_+ = \left [\frac{\alpha_s(M_W)}{\alpha_s(\mu)}\right ]^{\frac{6}{33-2f}},
\quad
C_- = \left [\frac{\alpha_s(M_W)}{\alpha_s(\mu)}\right ]^{\frac{-12}{33-2f}},\\
$$
where f is the number of flavors.     

Assuming that the energy release in the heavy quark decay is large, we can
perform the operator product expansion for the transition operator ${\cal T}$
in (\ref{1}). In this way we find a series of local operators with  
increasing dimension over the energy scale, wherein the contributions to
$\Gamma_{\Xi_{cc}^{(*)}}$ are suppressed by the increasing inverse powers of
the heavy quark masses. This formalism has already been applied to calculate
the total decay rates for the hadrons, containing a single heavy quark \cite{2}
(for the most early work, having used similar methods, see also
\cite{6,7}). Here we would like to stress that the expansion, applied in
this paper, is simultaneously in the powers of the inverse heavy quark mass
and the relative velocity of heavy quarks inside the hadron. Thus, the latter
points to the difference from the description of both the heavy-light mesons
(the expansion in powers of $\frac{\Lambda_{QCD}}{m_c}$) and the heavy-heavy
mesons \cite{8} (the expansion in powers of relative velocity of heavy quarks
inside the hadron, where one can apply the scaling rules of nonrelativistic QCD
\cite{4}).

In this work we will extend this approach to the treatment of baryons,
containing two heavy quarks. The operator product expansion applied has the
form:
\begin{equation}
{\cal T} = C_1(\mu)\bar cc + \frac{1}{m_c^2}C_2(\mu)\bar
cg\sigma_{\mu\nu}G^{\mu\nu}c
+ \frac{1}{m_c^3}O(1). \label{4}
\end{equation}

The leading contribution in OPE is determined by the operator $\bar cc$,
corresponding to the spectator decays of $c$-quarks. The use of
the equation of motion for the heavy quark fields allows one to eliminate some
redundant operators, so that no operators of dimension four contribute. There
is a single operator of dimension five, $Q_{GQ} = \bar Q g \sigma_{\mu\nu}
G^{\mu\nu} Q$. As we will show below, significant contributions come from the
operators of dimension six $Q_{2Q2q} = \bar Q\Gamma q\bar q\Gamma^{'}Q$,
representing the effects of Pauli interference and weak scattering for
$\Xi_{cc}^{++}$ and $\Xi_{cc}^{+}$, correspondingly. Furthermore, there are
also other operators of dimension six $Q_{61Q} = \bar Q \sigma_{\mu\nu}
\gamma_{l} D^{\mu}G^{\nu l}Q$ and $Q_{62Q} = \bar Q D_{\mu} G^{\mu\nu}
\Gamma_{\nu}Q$. In what follows, we do not calculate the corresponding
coefficient functions for the latter two operators, so that the expansion is
certainly complete up to the second order of $\frac {1}{m}$, only.

Further, the different contributions to OPE are given by the following 
\begin{eqnarray}
{\cal T}_{\Xi_{cc}^{++}} &=& {\cal T}_{35c} + {\cal T}_{6,PI},\nonumber\\
{\cal T}_{\Xi_{cc}^{+}} &=& {\cal T}_{35c} + {\cal T}_{6,WS}, \nonumber
\end{eqnarray}
where the first terms account for the operators of dimension three $O_{3Q}$ and
five $O_{GQ}$, the second terms correspond to the effects of Pauli interference
and weak scattering. The explicit formulae for these contributions have the
following form:
\begin{equation}
{\cal T}_{35c} = 2*(\Gamma_{c,spec}\bar cc - \frac{\Gamma_{0c}}{m_c^2}[(2 +
K_{0c})P_{1} + K_{2c}P_{2}]O_{Gc}), \label{5}
\end{equation}
where $\Gamma_{0c} = \frac{G_F^2m_c^2}{192{\pi}^3}$
and $K_{0c} = C_{-}^2 + 2C_{+}^2$,  $K_{2c} = 2(C_{+}^2 - C_{-}^2)$.
This expression has been derived in \cite{9} (see also \cite{10}), and it is
also discussed in \cite{2}. The phase space factors $P_i$ look like
\cite{2,11}:
$$
P_1 = (1-y)^4,\quad P_2 = (1-y)^3,
$$
where $y = \frac{m_s^2}{m_c^2}$.

$\Gamma_{c,spec}$ denotes the contribution to the total decay width
of the free decay for one of the two $c$-quarks, which is explicitly 
expressed below.

For the effects of Pauli interference and weak scattering, we find the
following formulae:
\begin{eqnarray}
{\cal T}_{PI} &=& -\frac{2G_F^2}{4\pi}m_c^2(1-\frac{m_u}{m_c})^2\nonumber\\
&& ([(\frac{(1-z_{-})^2}{2}- \frac{(1-z_{-})^3}{4}) 
(\bar c_i\gamma_{\alpha}(1-\gamma_5)c_i)(\bar
q_j\gamma^{\alpha}(1-\gamma_5)q_j) + \nonumber\\ 
&& (\frac{(1-z_{-})^2}{2} -
\frac{(1-z_{-})^3}{3})(\bar c_i\gamma_{\alpha}\gamma_5 c_i)(\bar
q_j\gamma^{\alpha}(1-\gamma_5)q_j)] \label{6}
\\&& [(C_{+} + C_{-})^2 + 
\frac{1}{3}(1-k^{\frac{1}{2}})(5C_{+}^2+C_{-}^2-6C_{-}C_{+})]+ \nonumber\\
&& [(\frac{(1-z_{-})^2}{2} - \frac{(1-z_{-})^3}{4})(\bar
c_i\gamma_{\alpha}(1-\gamma_5)c_j)(\bar q_j\gamma^{\alpha}(1-\gamma_5)c_i) +
\nonumber\\
&&  (\frac{(1-z_{-})^2}{2} - \frac{(1-z_{-})^3}{3})(\bar
c_i\gamma_{\alpha}\gamma_5c_j)(\bar
q_j\gamma^{\alpha}(1-\gamma_5)q_i)]k^{\frac{1}{2}}(5C_{+}^2+C_{-}^2-6C_{-}C_{+}
)),\nonumber\\
{\cal T}_{WS} &=& \frac{2G_F^2}{4\pi}p_{+}^2(1-z_{+})^2[(C_{+}^2 + C_{-}^2 +
 \frac{1}{3}(1 - k^{\frac{1}{2}})(C_{+}^2 - C_{-}^2))\nonumber\\
&&(\bar
c_i\gamma_{\alpha}(1
- \gamma_5)c_i)(\bar q_j\gamma^{\alpha}(1 - \gamma_5)q_j) + \label{7}\\
&& k^{\frac{1}{2}}(C_{+}^2 - C_{-}^2)(\bar c_i\gamma_{\alpha}(1 - \gamma_5)c_j)
(\bar q_j\gamma^{\alpha}(1 - \gamma_5)q_i)],\nonumber 
\end{eqnarray}
where $p_{+} = p_c + p_q$, $p_{-} = p_c - p_q$ and $z_{\pm} =
\frac{m_c^2}{p_{\pm}^2}$, $k=\alpha_s(\mu)/\alpha_s(m_c)$.

In the numerical estimates for the evolution of coefficients $C_{+}$ and
$C_{-}$, we have taken into account the threshold effects, connected to the
$b$-quark, as well as the threshold effects, related to the $c$-quark mass in
the Pauli interference and weak scattering.

In expression (\ref{5}), the scale $\mu$ is approximately equal to $m_c$. For
the Pauli interference and weak scattering, this scale was chosen in the way to
obtain an agreement between the experimental differences in the lifetimes of
$\Lambda_c$, $\Xi_c^{+}$ and $\Xi_c^{0}$-baryons and the theoretical
predictions, based on the effects, mentioned above. This problem is discussed
below. Anyway, the choice of these scales allows some variations, and a
complete answer to this question requires calculations in the next order of
perturbative theory.

The contribution of the leading operator $\bar cc$ corresponds to the imaginary
part of the diagram in Fig. 1, as it stands in expression (\ref{4}). The
coefficient of $\bar cc$ can be obtained in the usual way by matching of
the Fig. 1 diagram, corresponding to the leading term in expression (\ref{4}),
with the operator $\bar cc$. This coefficient is equivalent to the free quark
decay rate, and it is known in the next-to-leading logarithmic approximation of
QCD \cite{12,13,14,15,16}, including the strange quark mass effects in the
final state \cite{16}. To calculate the next-to-leading logarithmic effects,
the Wilson coefficients in the effective weak lagrangian are required at the
next-to-leading accuracy, and the single gluon exchange corrections to
the diagram in Fig.1 must be considered. In our numerical estimates we use the
expression for $\Gamma_{spec}$, including the next-to-leading order
corrections, $s$-quark mass effects in the final state, but we neglect the
Cabibbo-suppressed decay channels for the $c$-quark. The bulky explicit
expression for the spectator $c$-quark decay is placed in the Appendix.

Similarly, the contribution by $O_{GQ}$ is obtained, when an external gluon
line is attached to the inner quark lines in Fig. 1 in all possible ways. The 
corresponding coefficients are known in the leading logarithmic approximation.
Finally, the dimension six operators and their coefficients arise due to those
contributions, wherein one of the internal $u$ or $\bar d$ quark line is "cut"
in the diagram of Fig. 1. The resulting graphs are depicted in Figs. 2 and 3.
These contributions correspond to the effects of Pauli interference and weak
scattering. We have calculated the expressions for these effects with account
for both the $s$-quark mass in the final state and the logarithmic
renormalization of effective electroweak lagrangian at low energies.

Since the simultaneous account for the mass effects and low-energy logarithmic
renormalization of such contributions has been performed in this work for the
first time, we would like to discuss this question in some details.

The straightforward calculation of diagrams of Figs. 2 and 3 with the account
for the $s$-quark mass yields the following expressions:
\begin{eqnarray}
{\cal T}_{PI} &=&
-\frac{2G_F^2}{4\pi}p_{-}^2[(\frac{(1-z_{-})^3}{12}g^{\alpha\beta} + 
(\frac{(1-z_{-})^3}{2} -
\frac{(1-z_{-})^3}{3})\frac{p_{-}^{\alpha}p_{-}^{\beta}}{p_{-}^2})]\nonumber\\
&& [(C_{+} + C_{-})^2(\bar
c_i\gamma_{\alpha}(1-\gamma_5)c_i)(q_j\gamma_{\beta}(1-\gamma_5)q_j) + \\
&& (5C_{+}^2-6C_{+}C_{-}+C_{-}^2)(\bar
c_i\gamma_{\alpha}(1-\gamma_5)c_j)(q_j\gamma_{\beta}(1-\gamma_5)q_i)],
\nonumber\\
{\cal T}_{WS} &=&
\frac{2G_F^2m_c^2}{4\pi}p_{+}^2(1-z_{+})^2[(C_{+}^2+C_{-}^2)(\bar
c_i\gamma_{\alpha}(1-\gamma_5)c_i)(q_j\gamma_{\beta}(1-\gamma_5)q_j) +
\nonumber\\
&& (C_{+}^2-C_{-}^2)(\bar
c_i\gamma_{\alpha}(1-\gamma_5)c_j)(q_j\gamma_{\beta}(1-\gamma_5)q_i)].
\end{eqnarray}
For $p_{+}$ and $p_{-}$ we use their threshold values:
$$
p_{+}=p_c(1+\frac{m_q}{m_c}),\quad p_{-}=p_c(1-\frac{m_q}{m_c}),
$$
taking into account that the logarithmic renormalization of effective
low-energy lagrangian has the following form \cite{5,6}:
\begin{eqnarray}
L_{eff,log} &=&
\frac{G_f^2m_c^2}{2\pi}\{\frac{1}{2}[C_{+}^2+C_{-}^2+\frac{1}{3}(1-k^{\frac{1}{
2}})(C_{+}^2-C_{-}^2)](\bar c\Gamma_{\mu})(\bar d\Gamma^{\mu}d) +\nonumber\\
&& \frac{1}{2}(C_{+}^2-C_{-}^2)k^{\frac{1}{2}}(\bar c\Gamma_{\mu}d)(\bar
d\Gamma^{\mu} c) +
\frac{1}{3}(C_{+}^2-C_{-}^2)k^{\frac{1}{2}}(k^{\frac{-2}{9}}-1)(\bar
c\Gamma_{\mu}t^ac)j_{\mu}^a - \\
&& \frac{1}{8}[(C_{+}+C_{-})^2+\frac{1}{3}(1-k^{\frac{1}{2}})(5C_{+}^2+C_{-}^2-
6C_{+}C_{-})](\bar c\Gamma_{\mu}c+\frac{2}{3}\bar c\gamma_{\mu}\gamma_5c)(\bar
u\Gamma^{\mu}u)-\nonumber\\
&& \frac{1}{8}k^{\frac{1}{2}}(5C_{+}^2+C_{-}^2-6C_{+}C_{-})(\bar
c_i\Gamma_{\mu}c_k+\frac{2}{3}\bar c_i\gamma_{\mu}\gamma_5c_k)(\bar
u_k\Gamma^{\mu}u_i)-\nonumber\\
&& \frac{1}{8}[(C_{+}-C_{-})^2+\frac{1}{3}(1-k^{\frac{1}{2}})(
5C_{+}^2+C_{-}^2+6C_{+}C_{-})](\bar c\Gamma_{\mu}c+\frac{2}{3}\bar
c\gamma_{\mu}\gamma_5c)(\bar
s\Gamma^{\mu}s)-\nonumber\\
&& \frac{1}{8}k^{\frac{1}{2}}(5C_{+}^2+C_{-}^2+6C_{+}C_{-})(\bar
c_i\Gamma_{\mu}c_k+\frac{2}{3}\bar c_i\gamma_{\mu}\gamma_5c_k)(\bar
s_k\Gamma^{\mu}s_i)-\nonumber\\
&& \frac{1}{6}k^{\frac{1}{2}}(k^{\frac{-2}{9}}-1)(5C_{+}^2+
C_{-}^2)(\bar c\Gamma_{mu}t^ac+\frac{2}{3}\bar
c\gamma_{\mu}\gamma_5t^ac)j^{a\mu}\},
\nonumber
\end{eqnarray}
where $\Gamma_{\mu} = \gamma_{\mu}(1-\gamma_5)$,
$k=(\alpha_s(\mu)/\alpha_s(m_c))$ and $j_{\mu}^a = \bar u\gamma_{\mu}t^au +
\bar d\gamma_{\mu}t^ad + \bar s\gamma_{\mu}t^as$ is the color current of light
quarks ($t^a = \lambda^a/2$ being the color generators).  Having performed the
manipulations, we have obtained formulae (\ref{6}), (\ref{7}).

Here we would like to make a note, concerning the terms of effective
lagrangian, containing the color current of light quarks. In the analysis
below, we have omitted these terms, because they contribute in the 
lagrangian with the strength factor $k^{-\frac{2}{9}}-1$, whose numerical 
value is equal to $0.054$ (see below).

To calculate the  contribution of semileptonic modes to the total decay
width of $\Xi_{cc}^{(*)}$-baryons (we have taken into account the electron
and muon decay modes only) we use the following expressions \cite{10} (see also
\cite{16}):
\begin{eqnarray}
\Gamma_{sl} &=& 4\Gamma_c(\{1-8\rho+8\rho^3-\rho^4-12\rho^2\ln\rho\}
+\nonumber\\
&&
E_c\{5-24\rho+24\rho^2-8\rho^3+3\rho^4-12\rho^2\ln\rho\}+\\
&& K_c\{-6+32\rho-24\rho
^2-2\rho^4+24\rho^2\ln\rho\}+\nonumber\\
&& G_c\{-2+16\rho-16\rho^3+2\rho^4+24\rho^2\ln\rho\}),\nonumber 
\end{eqnarray}
where  $\Gamma_c = |V_{cs}|^2G_F^2\frac{m_c^5}{192\pi^3}$, $\rho
=\frac{m_s^2}{m_c^2}$. The quantities $E_c = K_c + G_c$,
 $K_c$ and  $G_c$ are given by the expressions:
\begin{eqnarray}
K_c &=& -\langle \Xi_{cc}^{(*)}(v)|\bar
c_v\frac{(iD)^2}{2m_c^2}c_v|\Xi_{cc}^{(*)}(v)\rangle ,\nonumber\\
G_c &=& \langle \Xi_{cc}^{(*)}(v)|\bar
c_v\frac{gG_{\alpha\beta}\sigma^{\alpha\beta}}{4m_c^2}c_v|\Xi_{cc}^{(*)}(v)
\rangle ,
\end{eqnarray}
where the spinor field $c_v$ in the effective heavy quark theory is defined by
the form:
\begin{equation}
c(x) = e^{-im_cv\cdot x}\Bigl[1+\frac{iD}{2m_c}\Bigr]c_v(x).
\end{equation}

Thus, we can see, that the evaluation of total lifetimes for the doubly
charmed baryons is reduced to the problem of estimation for the matrix
elements of operators, appearing in the above expressions, which is the topic
of next section.

\section{Evaluation of matrix elements.}

Let us calculate the matrix elements for the operators, obtained
as the result of OPE for the transitions under consideration. In general, it is
a complicated nonperturbative problem, but, as we will see below, in our
particular calculation we can get some reliable estimates for the matrix
elements of required operators.

Using the equation of motion for the heavy quarks, the local operator $\bar cc$
can be expanded in the following series over the powers of $\frac{1}{m_c}$:
\begin{eqnarray}
\langle \Xi_{cc}^{(*)}|\bar cc|\Xi_{cc}^{(*)}\rangle _{norm} = 1 -
\frac{\langle \Xi_{cc}^{(*)}|\bar
c[(i\vec D)^2-(\frac{i}{2}\sigma G)]c|\Xi_{cc}^{(*)}\rangle _{norm}}{2m_c^2} +
O(\frac{1}{m_c^3}).
\end{eqnarray}
Thus, this evaluation can be reduced to the calculation of matrix elements for
the following operators:  
$$
\bar c(i\vec D)^2c,\quad (\frac{i}{2})\bar c\sigma Gc,\quad \bar
c\gamma_{\alpha}(1-\gamma_5)c\bar q\gamma^{\alpha}(1-\gamma_5)q,\quad 
\bar c\gamma_{\alpha}\gamma_5c\bar q\gamma^{\alpha}(1-\gamma_5)q.
$$
The first operator corresponds to the time dilation, connected to the motion of
heavy quarks inside the hadron, the second is related to the spin interaction
of heavy quarks with the chromomagnetic field of light quark and the 
other heavy quark. Further, the third and fourth operators are the four-quark 
operators, representing the effects of Pauli interference and weak scattering.
  
In the system, containing the nonrelativistic heavy quark, the quark-antiquark
pairs with the same flavor can be produced with a negligible rate, since the
energy greater than $m_Q$ is required. In this situation, it is useful to
integrate out the small components of the heavy-quark spinor field and 
to present the result in terms of the two component spinor $\Psi_Q$. 
Following this approach,
we find that all contributions from virtualities greater than $\mu$, where
$m_c >  \mu >  m_cv_c$, can be explicitly taken into account in the
perturbative theory. This method is general and analogous to the effective
heavy quark theory. So,
\begin{eqnarray}
\bar cc &=& \Psi_c^{+}\Psi_c - \frac{1}{2m_c^2}\Psi_c^{+}(i\vec D)^2\Psi_c +
\frac{3}{8m_c^4}\Psi_c^{+}(i\vec D)^4\Psi_c -\nonumber\\
&& \frac{1}{2m_c^2}\Psi_c^{+}g\vec\sigma\vec B\Psi_c -
\frac{1}{4m_c^3}\Psi_c^{+}(\vec Dg\vec E)\Psi_c + ... \label{15}\\
\bar cg\sigma_{\mu\nu}G^{\mu\nu}c &=& -2\Psi_c^{+}g\vec\sigma\vec B\Psi_c -
\frac{1}{m_c}\Psi_c^{+}(\vec Dg\vec E)\Psi_c + ... \label{16}
\end{eqnarray}
In these expressions we have omitted the term $\Psi_c^{+}\vec\sigma (g\vec
E\times\vec D)\Psi_c$, corresponding to the spin-orbital interaction, because
it vanishes in the ground states of doubly charmed baryons. By definition, the
two-component spinor $\Psi_c$ has the same normalization as $Q$,
\begin{equation}
\int d^3x\Psi_c^{+}\Psi_c^{+} = \int d^3x Q^{+}Q.
\end{equation} 
Then, with the required accuracy, $\Psi_c$ can be expressed through the big
components of spinor $Q$
\begin{equation}
Q\equiv e^{-imt}\left(\phi\atop \chi\right)
\end{equation}
due to the following formula
\begin{equation}
\Psi_c = \left( 1 + \frac{(i\vec D)^2}{8m_c^2}\right)\phi.
\end{equation}
(this can be checked with the use of the equation of motion). Let us note that
the covariant derivative should be taken in the adjoint representation, when it
acts on the chromoelectric field,
\begin{equation}
(\vec D\vec E) = (\vec \partial T^a - gf^{abc}T^b\vec A^c)\vec E^a.
\end{equation}
Radiative corrections modify the coefficients of the chromomagnetic term 
$(\vec\sigma\vec B)$ and "Darwin" term in (\ref{15}). However, in the situation
at hand, these effects can be consistently neglected.

Now let us consider the significance of different contributions to the 
expansions in (\ref{15}) and (\ref{16}). 
Evaluating the contributions of chromomagnetic and
"Darwin" terms, we have to take into account the interaction of heavy quark
with the light quark as well as the interaction with the other heavy quark. 
In the first case, the procedure of calculation is analogous to that for the
heavy-light mesons. So, the "Darwin" term is suppressed by a factor of
$\frac{\Lambda_{QCD}}{m_c}$ in comparison with the chromomagnetic term, and,
thus, we neglect its contribution. In the second case, the analysis is
analogous to that for the heavy-heavy mesons, so that we can use the scaling
rules of nonrelativistic QCD \cite{4}. In this approach, the  contributions of
different operators can be estimated, using the following relations in Coulomb
gauge: 
$$
\Psi_c\sim (m_cv_c)^{\frac{3}{2}},\quad\vec D\sim m_cv_c,\quad gE\sim
m_c^2v_c^3,\quad gB\sim m_c^2v_c^4,\quad g\sim v_c^{\frac{1}{2}}.
$$
From these scaling rules for the heavy-heavy interaction, we can deduce that
the contribution of the "Darwin" term has the same order as that of 
chromomagnetic term.

Let us now start the calculation of matrix elements with the use of potential
models for the bound states of hadrons. While estimating the matrix element
value of the kinetic energy, we note, that the heavy quark kinetic energy
consists of two parts: the kinetic energy of the heavy quark motion inside the
diquark and the kinetic energy, related to the diquark motion inside the
hadron. According to the phenomenology of meson potential models, in the range
of average distances between the quarks: $0.1\; {\rm fm} < r < 1\; {\rm fm}$,
the average kinetic energy of quarks is constant and independent of both the
quark flavors, constituting meson, and the quantum numbers, describing the
excitations of the ground state. Therefore, we determine $T = m_dv_d^2/2 +
m_lv_l^2/2$ as the average kinetic energy of diquark and light quark, and $T/2
= m_{c1}v_{c1}^2/2 + m_{c2}v_{c2}^2/2$ as the average kinetic energy of heavy
quarks inside the diquark (the coefficient 1/2 takes into account the
antisymmetry of color wave function for the diquark). Finally,  we have the
following expression for the matrix element of the heavy quark kinetic energy:
\begin{equation}
\frac{\langle \Xi_{cc}^{(*)}|\Psi_c^{+}(i\vec
D)^2\Psi_c|\Xi_{cc}^{(*)}\rangle }{2M_{\Xi_{cc}^{(*)}}m_c^2}\simeq v_c^2\simeq
\frac{m_lT}{2m_c^2+m_cm_l}+\frac{T}{2m_c}.
\end{equation}
We use the value $T\simeq 0.4$ GeV, which results in $v_c^2 = 0.146$, where the
dominant contribution comes from the motion of heavy quarks inside the diquark.

Now we would like to estimate the matrix element of chromomagnetic operator,
corresponding to the interaction of heavy quarks with the chromomagnetic field
of the light quark. For this purpose, we will use the following definitions:
$O_{mag} = \sum_{i=1}^2 \frac{g_s}{4m_c}\bar c^i\sigma_{\mu\nu}G^{\mu\nu}c^i$
and $O_{mag}\sim \lambda (j(j+1) - s_d(s_d+1) - s_l(s_l+1))$, where $s_d$ is
the diquark spin (as was noticed by the authors earlier \cite{1}, there is only
the vector state of the $cc$-diquark in the ground state of such baryons), 
$s_l$ is
the light quark spin and $j$ is the total spin of the baryon. Since 
both $c$-quarks
additively contribute to the total decay width of baryons, we can use the
diquark picture and substitute for the sum of $c$-quark spins the diquark
spin. This leads to the parameterization for $O_{mag}$, as it is given above,
and, moreover, it allows us to relate the value of the matrix element for this
operator to the mass difference between the excited and ground state of
baryons:
\begin{equation}
O_{mag} = -\frac{2}{3}(M_{\Xi_{cc}^{(*)}}^{*}-M_{\Xi_{cc}^{(*)}}).
\end{equation}
The account for the interaction of heavy quarks inside the diquark leads to the
following expressions for the chromomagnetic and "Darwin" terms:
\begin{eqnarray}
\frac{\langle \Xi_{cc}^{(*)}|\Psi_c^{+}g\vec\sigma\cdot \vec
B\Psi_c|\Xi_{cc}^{(*)}\rangle }{2M_{\Xi_{cc}^{(*)}}} &=&
\frac{2}{9}g^2\frac{|\Psi
(0)|^2}{m_c}, \\
\frac{\langle \Xi_{cc}^{(*)}|\Psi_c^{+}(\vec D\cdot g\vec
E)\Psi_c|\Xi_{cc}^{(*)}\rangle }{2M_{\Xi_{cc}^{(*)}}} &=& \frac{2}{3}g^2|\Psi
(0)|^2.
\end{eqnarray}
where $\Psi (0)$ is the diquark wave function at the origin.

Collecting the results given above, we find the matrix elements of operators
(\ref{15}) and (\ref{16}):
\begin{eqnarray}
\frac{\langle \Xi_{cc}^{(*)}|\bar cc|\Xi_{cc}^{(*)}\rangle
}{2M_{\Xi_{cc}^{(*)}}} &=& 1 -
\frac{1}{2}v_c^2 - \frac{1}{3}\frac{M_{\Xi_{cc}^{(*)}}-M_{\Xi_{cc}^{(*)}}}{m_c}
- \frac{g^2}{9m_c^3}|\Psi (0)|^2 - \nonumber\\
&& \frac{1}{6m_c^3}g^2|\Psi (0)|^2 + ... \nonumber\\
&\approx& 1 - 0.074 -0.004 - 0.003 -
0.005+\ldots
\end{eqnarray}

We can see that the largest contribution to the decrease of the
decay width comes from the time dilation, connected to the motion of heavy
quarks inside the baryon. For the matrix element of the operator $\bar
cg\sigma_{\mu\nu}G^{\mu\nu}c$, we get:
\begin{eqnarray}
\frac{\langle \Xi_{cc}^{(*)}|\bar
cg\sigma_{\mu\nu}G^{\mu\nu}c|\Xi_{cc}^{(*)}\rangle }{2M_{\Xi_{cc}^{(*)}}} &=&
-\frac{4}{3}\frac{(M_{\Xi_{cc}^{(*)}}^{*} - 
M_{\Xi_{cc}^{(*)}})}{m_c} \nonumber\\
&& - \frac{4g^2}{9m_c^3}|\Psi (0)|^2 - \frac{g^2}{3m_c^3}|\Psi (0)|^2.
\end{eqnarray}
Now let us continue with the calculation of the matrix elements for 
the four-quark
operators, corresponding to the effects of Pauli interference and weak
scattering. The straightforward calculation in the framework of nonrelativistic
QCD gives
\begin{equation}
(\bar c\gamma_{\mu}(1-\gamma_5)c)(\bar q\gamma^{\mu}(1-\gamma_5)q) =
2m_cV^{-1}(1-4S_cS_q),
\end{equation}
\begin{equation}
(\bar c\gamma_{\mu}\gamma_5c)(\bar q\gamma^{\mu}(1-\gamma_5)q) = -4S_cS_q\cdot
2m_cV^{-1},
\end{equation}
where $V^{-1} = |\Psi_1(0)|^2$, and $\Psi_1(0)$ is the light quark
wave function at the origin of two $c$-quarks. We suppose, that 
$|\Psi_1(0)|$ has the same value as that in the $D$-meson. So, we find:
\begin{equation}
|\Psi_1(0)|^2\approx\frac{f_D^2m_D^2}{12m_c}.
\end{equation}
Then, again remembering that both $c$-quarks additively contribute to
the total decay width and using the diquark picture, we can substitute
for $S_{c_1} + S_{c_2}$ the $S_d$, where $S_d$ is the diquark spin. Thus, we
have
\begin{eqnarray}
\langle \Xi_{cc}^{(*)}|(\bar c\gamma_{\mu}(1-\gamma_5)c)(\bar
q\gamma^{\mu}(1-\gamma_5)q)|\Xi_{cc}^{(*)}\rangle  &=& 10m_c\cdot
|\Psi_1(0)|^2,\\
\langle \Xi_{cc}^{(*)}|(\bar c\gamma_{\mu}\gamma_5c)(\bar
q\gamma^{\mu}(1-\gamma_5)q)|\Xi_cc^{(*)}\rangle  &=& 8m_c\cdot
|\Psi_1(0)|^2.
\end{eqnarray}
The color antisymmetry of the baryon wave function results in relations
between the matrix elements of operators with the different sums over the
color indexes:
$$\langle \Xi_{cc}^{(*)}|(\bar
c_iT_{\mu}c_k)(\bar q_k\gamma^{\mu}(1-\gamma_5)q_i|\Xi_{cc}^{(*)}\rangle  =
-\langle \Xi_{cc}^{(*)}|(\bar cT_{\mu}c)(\bar
q\gamma^{\mu}(1-\gamma_5)q|\Xi_{cc}^{(*)}\rangle ,
$$
where $T_{\mu}$ is any spinor structure. Thus, we completely derive the
expressions for the evaluation of the required matrix elements.

\section{Numerical estimates}

Now we are ready to collect the contributions, described above, and to estimate
the total lifetimes of baryons $\Xi_{cc}^{++}$ and $\Xi_{cc}^{+}$. For the
beginning, we list the values of parameters, which we have used in our
calculations, and give some comments on their choice.
$$
\begin{array}{rclrclrcl}
m_c &=& 1.6~~GeV, & m_s &=& 0.45~~GeV, &|V_{cs}| &=& 0.9745,\\
M_{\Xi_{cc}^{++}} &=& 3.56~~GeV, &M_{\Xi_{cc}^{+}} &=& 3.56~~GeV,&
M_{\Xi_{cc}^{(*)}}^{*} - M_{\Xi_{cc}^{(*)}} &=& 0.1~~GeV,\\ 
T &=& 0.4~~GeV,& |\Psi (0)| &=& 0.17~~GeV^{\frac{3}{2}},& m_l &=&
0.30~~GeV.
\end{array}
$$
For the parameters $M_{\Xi_{cc}^{++}}$, $M_{\Xi_{cc}^{+}}$ and
$M_{\Xi_{cc}^{(*)}}^{*} - M_{\Xi_{cc}^{(*)}}$ we use the mean values, given
in the literature. Their evaluation has been also performed by the authors 
in the
potential model for the doubly charmed baryons with the Buchm\" uller-Tye
potential, and also in papers \cite{20,21,22,23}. For $f_D$ we use the
value, given in \cite{6,24} and for $T$ we take it from \cite{25}.  
The mass $m_c$ corresponds to the pole mass of the $c$-quark. For its 
determination
we have used a fit of theoretical predictions for the lifetimes and
semileptonic width of the $D^0$-meson from the experimental data. 
This choice of
$c$-quark mass seems effectively to include unknown contributions of higher
orders in perturbative QCD to the total decay width of baryons under
consideration.  

The renormalization scale $\mu$ is chosen in the following way:
$\mu_1 = m_c$ in the estimate of Wilson coefficients $C$ for the effective
four-fermion weak lagrangian with the $c$-quarks at low energies and $\mu_2 =
1.2~~GeV$ for the Pauli interference and weak scattering ($k$-factor). The
latter value of renormalization scale has been obtained from the fit of
theoretical predictions for the lifetimes differences of baryons
$\Lambda_c$, $\Xi_c^{+}$, $\Xi_c^{0}$ over the experimental data.
Here we would like to note, that the theoretical approximations used in 
\cite{5}
include the effect of logarithmic renormalization and do not take into account
the mass effects, related to the $s$-quark in the final state. For the
corresponding contributions to the decay widths of baryons with the different 
quark content we have: 
\begin{eqnarray}
\triangle\Gamma_{nl}(\Lambda_c) &=& c_d\langle O_d\rangle _{\Lambda_c} +
c_u\langle O_u\rangle _{\Lambda_c}, \nonumber\\
\triangle\Gamma_{nl}(\Xi_c^{+}) &=& c_s\langle O_s\rangle _{\Xi_c^{+}} +
c_u\langle O_u\rangle _{\Xi_c^{+}},\\
\triangle\Gamma_{nl}(\Xi_c^{0}) &=& c_d\langle O_d\rangle _{\Xi_c^{0}} +
c_s\langle O_s\rangle _{\Xi_c^{0}}, \nonumber
\end{eqnarray}
where $\langle O_q\rangle _{X_c} = \langle X_c|O_q|X_c\rangle $, $O_q = (\bar
c\gamma_{\mu}c)(\bar
q\gamma^{\mu}q)$ and $q = u, d, s$. The coefficients $c_q(\mu)$ are equal to:
\begin{eqnarray}
c_d &=& \frac{G_f^2m_c^2}{4\pi}[C_{+}^2 + C_{-}^2 +
\frac{1}{3}(4k^{\frac{1}{2}} - 1)(C_{-}^2 - C_{+}^2)], \nonumber\\
c_u &=& -\frac{G_f^2m_c^2}{16\pi}[(C_{+} + C_{-})^2 + \frac{1}{3}(1 -
4k^{\frac{1}{2}})(5C_{+}^2 + C_{-}^2 - 6C_{+}C_{-})], \\
c_s &=& -\frac{G_f^2m_c^2}{16\pi}[(C_{+} - C_{-})^2 + \frac{1}{3}(1 -
4k^{\frac{1}{2}})(5C_{+}^2 + C_{-}^2 + 6C_{+}C_{-})].\nonumber
\end{eqnarray}
We use the spin averaged value of the $D$-meson mass for the estimation of 
the effective light quark mass $m_l$ as it stands below:
\begin{equation}
m_D = m_c + m_l + \frac{T\cdot m_l}{m_c+m_l} \approx 1.98~~GeV.
\end{equation}
The $s$-quark mass can be written down as:
\begin{equation}
m_s = m_l + 0.15\; GeV.
\end{equation}
As we have already mentioned, the spectator decay width of $c$-quark
$\Gamma_{c,spec}$ is known in the next-to-leading order of perturbative QCD
\cite{12,13,14,15,16}. The most complete calculation, including the mass
effects, connected to the $s$-quark in the final state, is given in \cite{16}.
In the present work we have used the latter result for the calculation of
the spectator contribution to the total decay width of doubly charmed baryons.
In the calculation of the semileptonic decay width, we neglect the electron 
and muon masses in the final state. Moreover, we neglect the 
$\tau$-lepton mode.

Now, let us proceed with the numerical analysis of contributions by the
different decay modes into the total decay width. In table 1 we have listed the
results for the fixed values of parameters, described above. From this table
one can see the significance of effects caused by both the Pauli interference
and the weak scattering in the decays of doubly charmed baryons. The Pauli
interference gives the negative correction about $36\%$ for the
$\Xi_{cc}^{++}$-baryons, and the weak scattering increases the total width by
$144\%$ for $\Xi_{cc}^{+}$. As it has been already noted in the Introduction, 
these effects take place differently in the baryons, and, thus, 
they enhance the difference of lifetimes for these hadrons. 

It is worth here to recall that the lifetime difference of $D^{+}$ and
$D^{0}$-mesons is generally explained by the Pauli interference of $c$-quark
decay products with the antiquark in the initial state, while in the current
consideration, we see the dominant contribution of weak scattering. This could
not be surprising, because under a more detailed consideration we will find, 
that the formula for the Pauli interference operator  for the $D$-meson 
coincides with that for the weak scattering in the case of baryons, 
containing, at least, a single $c$-quark. 

Finally, collecting the different contributions for the total lifetimes of
doubly charmed baryons, we obtain the following values:
$$\tau_{\Xi_{cc}^{++}} = 0.43\; ps,\;\;\;
\tau_{\Xi_{cc}^{+}} = 0.12\; ps.
$$
Rather broad variations of both the $c$-quark mass in the range of $1.6 -
1.65~~GeV$ and the mass difference for the strange and ordinary light quarks in
(35) in the range of $0.15 - 0.2~~GeV$, lead to the uncertainties  in the
lifetimes:
$\delta\tau_{\Xi_{cc}^{++}} = \pm 0.1~~ps$,
$\delta\tau_{\Xi_{cc}^{+}} = \pm 0.01~~ps.$

\section{Conclusion}

In this work we have performed a detailed investigation on the lifetimes of
doubly charmed baryons $\Xi_{cc}^{++}$, $\Xi_{cc}^{+}$ on the basis of the
operator product expansion for the transition currents. For the first time, we
have presented the formulae with the simultaneous account for both the
mass effects and low-energy logarithmic renormalization for the contributions
to the total decay width of baryons, containing heavy quarks,
as it is caused by the effects of Pauli interference and weak scattering. 
The usage of the diquark picture has allowed us to evaluate the matrix 
elements of operators derived. Further, we have discussed the 
procedure to choose the
values of parameters for the total lifetimes of these baryons.  The obtained
results show the significant role of both the Pauli interference and the weak
scattering. 

In conclusion, the authors would like to express their gratitude 
to Prof. S.S.Gershtein for usefull discussions and, especially, to 
Prof. M.B.Voloshin for clear explanations to some questions appeared during
a walking along this work. We want also to thank for the hospitality of
DESY-Theory group, where a part of this work was done.

This work is in part supported by the Russian Foundation for Basic Research,
grants 96-02-18216 and 96-15-96575.

\section*{Appendix}

In this appendix we present the explicit formulae \cite{16} for the spectator
decay of the $c$-quark in next-to-leading order of the perturbation theory
with the account for the mass effects, related to the $s$-quark in the final
state.

The coefficients $C_{+}$ and $C_{-}$ in the effective lagrangian with the
account for the next-to-leading order in perturbative QCD acquire the
additional multiplicative factors:
$$
F_{\pm}(\mu) = 1 + \frac{\alpha_s(m_W) -
\alpha_s(\mu)}{4\pi}\frac{\gamma_{\pm}^{(0)}}{2\beta_0}\left(\frac{\gamma_{\pm}
^{(1)}}{\gamma_{\pm}^{(0)}} - \frac{\beta_1}{\beta_0}\right) +
\frac{\alpha_s(m_W)}{4\pi}B_{\pm},
$$
where $\gamma_{\pm}^{(i)}$ is the coefficients of anomalous dimensions for the
operators $O_{\pm}$:
$$
\gamma_{\pm} = \gamma_{\pm}^{(0)}\frac{\alpha_s}{4\pi} +
\gamma_{\pm}^{(1)}\left(\frac{\alpha_s}{4\pi}\right)^2 + O(\alpha_s^3),
$$
with 
$$
\gamma_{+}^{(0)} = 4,\quad \gamma_{-}^{(0)} = -8,\quad \gamma_{+}^{(1)} = -7 +
\frac{4}{9}n_f,\quad \gamma_{-}^{(1)} = -14 - \frac{8}{9}n_f,
$$
in the naive dimensional regularization (NDR) with the anticommutating
$\gamma_5$, and $n_f$ is a number of flavours taken into account. $\beta_i$ is
the initial two coefficients of QCD $\beta$-function,
\begin{eqnarray}
&&
\beta = -g_s\{\beta_0\frac{\alpha_s}{4\pi} +
\beta_1\left(\frac{\alpha_s}{4\pi}\right)^2 + O(\alpha_s^3)\},\nonumber\\
&&
\beta_0 = 11 - \frac{2}{3}n_f,\quad \beta_1 = 102 - \frac{38}{3}n_f.\nonumber
\end{eqnarray}
The coefficients $B_{\pm}$ are written down in accordance to the requirement of
agreement between the effective lagrangian, evaluated at the scale $\mu = m_W$,
and the Standard Model one up to terms of the order of $\alpha_s^2(m_W)$:
$$
B_{\pm} = \pm B\frac{N_c\mp 1}{2N_c},
$$
where $N_c = 3$ is the number of colors. In the NDR scheme for $B$, we find 
$B = 11$.

Using the effective lagrangian in the next-to-leading order of perturbative QCD
and calculating the one-gluon corrections, we get the following expression for
the spectator $c$-quark decay:
\begin{eqnarray}
\Gamma(c\to su\bar d) &=& \Gamma_0[2C_{+}^2(\mu) + C_{-}^2(\mu) +
\frac{\alpha_s(m_W) - \alpha_s(\mu)}{2\pi}\{2C_{+}^2(\mu)R_{+} +
C_{-}^2(\mu)R_{-}\} + \nonumber\\
&& \frac{\alpha_s(\mu)}{2\pi}\{2C_{+}^2(\mu)B_{+} +
C_{-}^2(\mu)B_{-}\} + \nonumber\\
&& \frac{3}{4}\{C_{+}(\mu) +
C_{-}(\mu)\}^2\frac{2}{3}\frac{\alpha_s(\mu)}{\pi}\{G_a + G_b\} +
\nonumber\\
&& \frac{3}{4}\{C_{+}(\mu) -
C_{-}(\mu)\}^2\frac{2}{3}\frac{\alpha_s(\mu)}{\pi}\{G_c + G_d\} +
\\
&& \frac{1}{2}\{C_{+}^2(\mu) -
C_{+}^2(\mu)\}\frac{2}{3}\frac{\alpha_s(\mu)}{\pi}\{G_a+G_b+Ge\}],
\nonumber
\end{eqnarray}
where
\begin{eqnarray}
\Gamma_0 &=& \frac{G_f^2m_c^5}{192\pi^3}|V_{cs}|^2f_1(m_s^2/m_c^2),\nonumber\\
f_1(a) &=& 1 - 8a + 8a^3 - a^4 -12a^2\ln a,\nonumber
\end{eqnarray}
and
$$
R_{\pm} = B_{\pm} +
\frac{\gamma_{\pm}^{(0)}}{2\beta_0}\left(\frac{\gamma_{\pm}^{(1)}}{\gamma_{\pm}
^{(0)}} - \frac{\beta_1}{\beta_0}\right).
$$
For $G_a, G_b, G_c, G_d$ and $G_e$ we have found: 
\begin{eqnarray}
(G_a + G_b)f_1(a) &=& \frac{31}{4} - \pi^2 -a[80-\ln
a]+32a^{3/2}\pi^2\nonumber\\
&& -a^2[273 + 16\pi^2 - 18\ln a + 36\ln^2a] + 32a^{5/2}\pi^2\nonumber\\
&& -\frac{8}{9}a^3[118 - 57\ln a] + O(a^{7/2}),\\
(G_c + G_d)f_1(a) &=& \frac{31}{4} - \pi^2 - 8a[10 - \pi^2 + 3\ln a] - a^2[117
- 24\pi^2 +\nonumber\\
&& (30-8\pi)\ln a + 36\ln^2a] - \frac{4}{3}a^3[79+2\pi^2\nonumber\\
&& -62\ln a + 6\ln^2a] + O(a^4),\\
(G_a+G_b+G_e+B)f_1(a)
&=& (6\ln\frac{m_c^2}{\mu^2}+11)f_1(a) - \frac{51}{4} -\pi^2 + 8a[21-\pi^2
-3\ln a]\nonumber\\
&& + 32a^{3/2}\pi^2 - a^2[111 + 40\pi^2 - 258\ln a + 36\ln^2a]\nonumber\\
&& + 32a^{5/2}\pi^2 - \frac{4}{9}a^3[305 + 18\pi^2 + 30\ln a -
54\ln^2a]\nonumber\\
&& + O(a^{7/2}).
\end{eqnarray}
These approximations can be used in the range of $a$ values:
$a < 0.15$, where $a =
(\frac{m_s}{m_c})^2$, which, indeed, takes place in the calculations under
consideration.

\newpage
\begin{table}[t]
\begin{tabular}{|p{30mm}|r|p{30mm}|p{30mm}|}
\hline
Mode or decay mechanism & Width, $ps^{-1}$ & Contribution in $\%$
($\Xi_{cc}^{++}$) & Contribution in $\%$ ($\Xi_{cc}^{+}$) \\
\hline
$c_{spec}\to s\bar du$ & $2.894$~~ & ~~~124 & ~~~32\\
\hline
$c\to se^{+}\nu$ & $0.380$~~ & ~~~~16 & ~~~~4\\
\hline
PI & $-1.317$~~ & ~~~-56 & ~~~~--\\ 
\hline
WS & $5.254$~~ & ~~~~-- & ~~~59\\
\hline
 $\Gamma_{\Xi_{cc}^{++}}$ & $2.337$~~ & ~~~100 &~~~--\\
\hline
$\Gamma_{\Xi_{cc}^{+}}$ & $8.909$~~ & ~~~--  &~~100\\
\hline
\end{tabular}
\caption{The contributions of different modes to the total decay width
of doubly charmed baryons.}
\end{table}

\newpage
\setlength{\unitlength}{1mm}
\begin{figure}[p]
\vspace*{-9mm}
\begin{center}
\begin{picture}(400.,30.)
\hspace*{3cm}
\epsfxsize=9cm \epsfbox{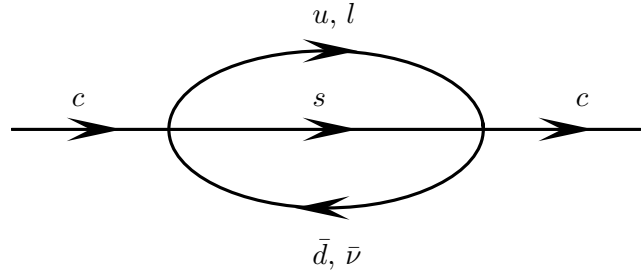}
\put(-82,24){$c$}
\put(-15,24){$c$}
\put(-50,24){$s$}
\put(-50,35){$u$, $l$}
\put(-50,3){$\bar d$, $\bar \nu$}

\end{picture}
\end{center}
\vspace*{-8mm}
\caption{The spectator contribution to the total decay width of doubly charmed
baryons.}
\label{fig1}
\end{figure}

\begin{figure}[p]
\vspace*{-9mm}
\begin{center}
\begin{picture}(400.,30.)
\hspace*{3cm}
\epsfxsize=9cm \epsfbox{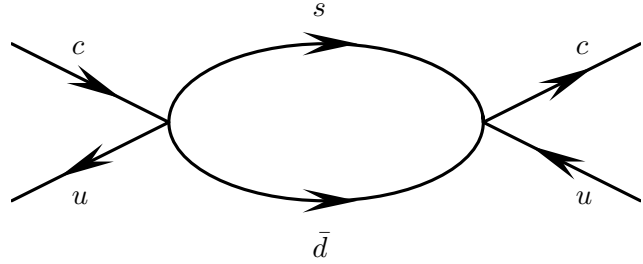}
\put(-82,30){$c$}
\put(-82,10){$u$}
\put(-15,30){$c$}
\put(-15,10){$u$}
\put(-50,35){$s$}
\put(-50,3){$\bar d$}

\end{picture}
\end{center}
\vspace*{-8mm}
\caption{The Pauli interference of $c$-quark decay products
with the valence quark in the initial state for the $\Xi_{cc}^{++}$-baryon.}
\label{fig2}
\end{figure}

\begin{figure}[p]
\vspace*{-9mm}
\begin{center}
\begin{picture}(400.,30.)
\hspace*{3cm}
\epsfxsize=9cm \epsfbox{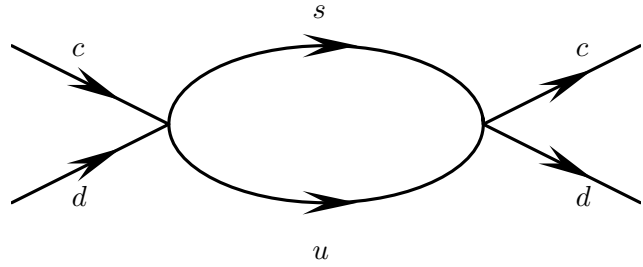}
\put(-82,30){$c$}
\put(-82,10){$d$}
\put(-15,30){$c$}
\put(-15,10){$d$}
\put(-50,35){$s$}
\put(-50,3){$u$}

\end{picture}
\end{center}
\vspace*{-8mm}
\caption{The weak scattering of the valence quarks in the initial
state for the $\Xi_{cc}^{+}$-baryon.}
\label{fig3}
\end{figure}

\end{document}